\documentclass[10pt,leqno,a4paper,oneside]{amsart}
\usepackage[utf8]{inputenc}
\usepackage{graphicx}
\usepackage{indentfirst,csquotes}
\usepackage{amssymb,amsthm,amsmath}
\usepackage{xcolor,paralist,hyperref,titlesec,fancyhdr,etoolbox}
\usepackage{lipsum}
\usepackage{caption}
\usepackage[a4paper, margin=1.5cm]{geometry}

\titleformat{\section}
  {\normalfont\Large\bfseries\centering} 
  {\thesection.}                         
  {1em}                                  
  {}                                     

\titleformat{\subsection}
  {\normalfont\large\bfseries}           
  {\thesubsection.}                      
  {1em}                                  
  {}

\titlespacing*{\section}{0pt}{3.5ex plus 1ex minus .2ex}{2.3ex plus .2ex}
\titlespacing*{\subsection}{0pt}{3ex plus 1ex minus .2ex}{1.5ex plus .2ex}

\hypersetup{ colorlinks=true, linkcolor=blue, filecolor=black, urlcolor=blue }

\title{Acoustic Guided Waves in MoS$_{2}$ thin flakes} 

\author{
    Aversa Martín$^{1,2}$, Nicolás Roqueiro$^{1,2}$, Camila Borrazás$^{1,3,4}$, Juan Ignacio Sangiorgio$^{1,2}$, Hilario D. Boggiano$^{1}$, Juan Bonaparte$^{5}$, Andrés Di Donato$^{5}$, María Cecilia Fuertes$^{3,4}$, Andrea V. Bragas$^{1,2,*}$ and Gustavo Grinblat$^{1,2,\dagger}$
}

\address{}

\begin{document}

\maketitle

\vspace{-0.8cm} 
\begin{flushleft}
    {\footnotesize \textnormal{
        \hspace{-0.09cm}$^1$ Universidad de Buenos Aires, Facultad de Ciencias Exactas y Naturales, Departamento de Física, Argentina. \\
        $^2$ CONICET-UBA, Instituto de Física de Buenos Aires (IFIBA), Buenos Aires, Argentina. \\
        $^3$ CNEA-CONICET Gerencia Química e INN. \\
        $^4$ UNSAM-CNEA, Instituto Sábato. \\
        $^5$ CAC-CNEA, Departamento de Micro y Nano Tecnología. \\
        \vspace{0.1cm}
        $^*$ Email: \href{mailto:bragas@df.uba.ar}{bragas@df.uba.ar} \quad
        $^\dagger$ Email: \href{mailto:grinblat@df.uba.ar}{grinblat@df.uba.ar}
    }}
\end{flushleft}

\vspace{0.5cm}

\begin{quotation}
\noindent 
Guided acoustic waves in two-dimensional materials are a key channel for energy transport and dissipation, yet their generation and propagation in transition metal dichalcogenides remain poorly understood. Here, we employ in situ and spatially decoupled ultrafast optical pump–probe techniques to investigate guided waves in MoS$_2$ flakes with thicknesses between 90 and 410 nm. We observe a propagating acoustic excitation with a constant velocity of (6.7~$\pm$~0.8)~km~s$^{-1}$ , independent of thickness. Finite element simulations and symmetry analysis reveal that these vibrations deviate from the classical Lamb wave model and are better described as a superposition of decoupled longitudinal and shear modes. We show that their optical detectability is governed by the Poisson effect: longitudinal components modulate the flake thickness and generate a measurable signal, whereas shear motion remains largely optically invisible. An intrinsic attenuation length of $\chi \sim$ 3.3~\textmu{}m indicates that dissipation is dominated by material-specific mechanisms rather than geometric spreading. Finally, we demonstrate remote excitation across a nanometric step, enabling acoustic generation in optically inaccessible regions. These results provide a foundation for nanoscale phononic circuits and engineered in-plane energy transport in 2D-based optomechanical and quantum acoustic devices.
\end{quotation}

\vspace{0.8cm}


\section{Introduction}
Transition metal dichalcogenides (TMDs) are a family of materials whose study has increased significantly over the last decade. This surge in interest is driven by the search for two dimensional semiconductors, a gap not fulfilled by graphene due to its semimetallic nature. These materials can be synthesized at the nanoscale using various methods, including mechanical exfoliation \cite{HUANG2020}, chemical vapor deposition \cite{WANG2020}, and molecular beam epitaxy \cite{FU2017}. TMDs exhibit remarkable properties, such as strong spin-valley coupling \cite{XIAO2012}, high in-plane carrier mobility \cite{Kaasbjerg2013}, and a robust light-emission capacity \cite{Splendiani2010}. The latter arises from the indirect-to-direct bandgap transition that these semiconductors undergo as they are thinned down to a monolayer. Furthermore, the high degree of tunability of these properties \cite{Conley2013}\cite{Kang2017}\cite{Castellanos2013} enhances their potential for use in electronics \cite{Radisavljevic2011}, optoelectronics \cite{WANG2012}\cite{Huang2016}, and valleytronics applications \cite{Mak2012}\cite{Liu2019}, among others \cite{YANG2023}\cite{Manzeli2017}\cite{CHOI2017}\cite{Nam2015}.

TMDs are also promising materials for controlling and harvesting acoustic vibrations at the nanoscale. In particular, TMD-based acoustic nanocavities (ANCs) enable the confinement of coherent phonons at the nanoscale and provide a platform for exploring high-frequency optomechanical interactions. Their relevance spans emerging applications in quantum computing \cite{Qiao2023,Chu2017} and optomechanics \cite{Fang2016}.
Current efforts aim to push operation into the GHz and THz regimes while enhancing phonon lifetimes, thereby improving device performance and sensitivity.

One class of ANCs exploits the out-of-plane breathing mode oscillations of van der Waals materials, whose frequencies depend on the number of layers and have been extensively studied. The excitation and detection of these modes are typically performed using ultrafast optical techniques \cite{Jeong2016,Boschetto2013}. State-of-the-art studies on TMDs report ANCs with quality factors as high as 10$^{14}$ in the 50–600~GHz range \cite{Zalalutdinov2021}, explore applications based on higher harmonics \cite{AVERSA2026,CARR2024}, and present detailed investigations of energy dissipation mechanisms \cite{AVERSA2026, Soubelet2019}.

In addition to breathing modes, guided acoustic wave modes constitute another excitation channel accessible in TMD-based ANCs via ultrafast optical interactions \cite{Zalalutdinov2021,AVERSA2026}. These guided modes include Lamb waves (LW), Rayleigh waves (RW), transverse horizontal waves (TH), and surface acoustic waves (SAWs), as described in classical elasticity theory for plates and surfaces \cite{Royer2000,Zhongqing2009}. Such acoustic modes are widely used in applications ranging from structural health monitoring and nondestructive testing in engineering \cite{Yu2008,Qing2019,Miao2021,Mei2019} to shear-wave elastography in medical diagnostics \cite{Taljanovic2017,Hyun2017}.

To date, devices exploiting these acoustic modes have primarily relied on piezoelectric substrates such as LiNbO$_3$ \cite{Preciado2015,Zhao2023}, GaAs \cite{Polimeno2024}, and related systems \cite{Kalameitsev2019,Mohajerani2025}, which generate acoustic waves that subsequently couple to the TMD and modulate its properties \cite{Reddeppa2025,Sheng2020}. Consequently, a comprehensive understanding of the generation mechanisms and intrinsic properties of guided acoustic waves directly within TMDs remains lacking.

In this work, we employ a spatially decoupled pump–probe technique to investigate guided acoustic wave modes in MoS$_2$ flakes on a silica substrate. We measure the propagation velocities of these modes and compare the experimental results with finite element method (FEM) simulations, revealing the nature of the underlying elastic modes. These results provide direct insight into the intrinsic generation and propagation of guided acoustic waves in TMDs, establishing a foundation for the engineering of guided-wave modes in two-dimensional materials for nanoscale phononics, optomechanics, and quantum acoustic devices.

\section{Results and Discussion}

\begin{figure}
  \centering
  \includegraphics{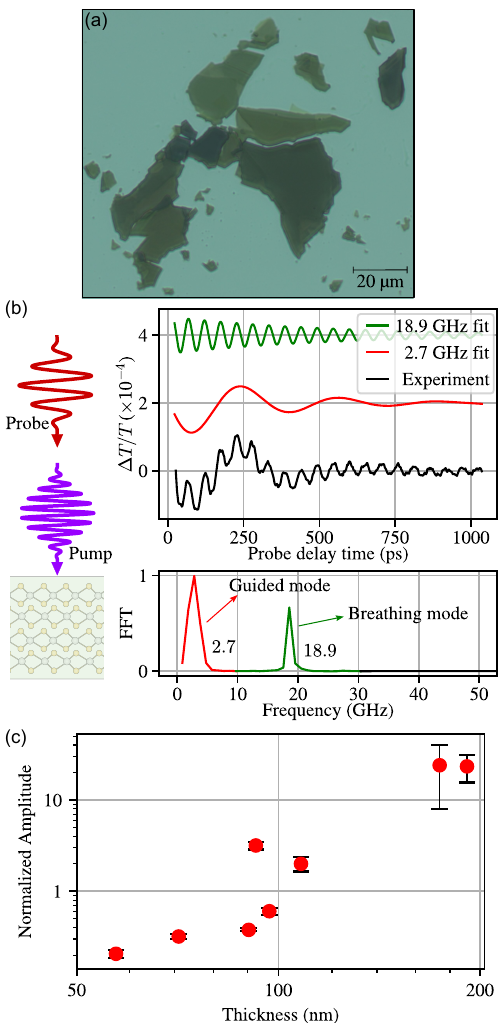}
  \caption{Basic characterization of guided acoustic waves in MoS$_2$. (a) Optical transmission microscopy image of mechanically exfoliated MoS$_2$ flakes on a SiO$_2$ substrate. (b) Left: Schematic of the in situ pump–probe configuration: a 400~nm wavelength pump pulse excites acoustic vibrations in the MoS$_2$ flake, while an 800~nm wavelength probe pulse monitors the resulting changes in optical transmission induced by the lattice oscillations. Both spots are spatially overlapped on the sample surface. Right: Representative in situ differential transmission signal after thermal background subtraction (top, black solid line) and its corresponding FFT (bottom) for a 108~nm thick flake. The spectrum reveals the fundamental breathing mode at 18.9~GHz (green) and a lower-frequency guided wave component at 2.7~GHz (red). The red and green solid lines in the top panel represents a damped sine fit of the guided and breathing mode, respectively. (c) Normalized amplitude of the guided wave mode relative to the fundamental breathing mode for all in situ measurements as a function of flake thickness.
}
  \label{fig1}
\end{figure}

\begin{figure}
  \centering
  \includegraphics{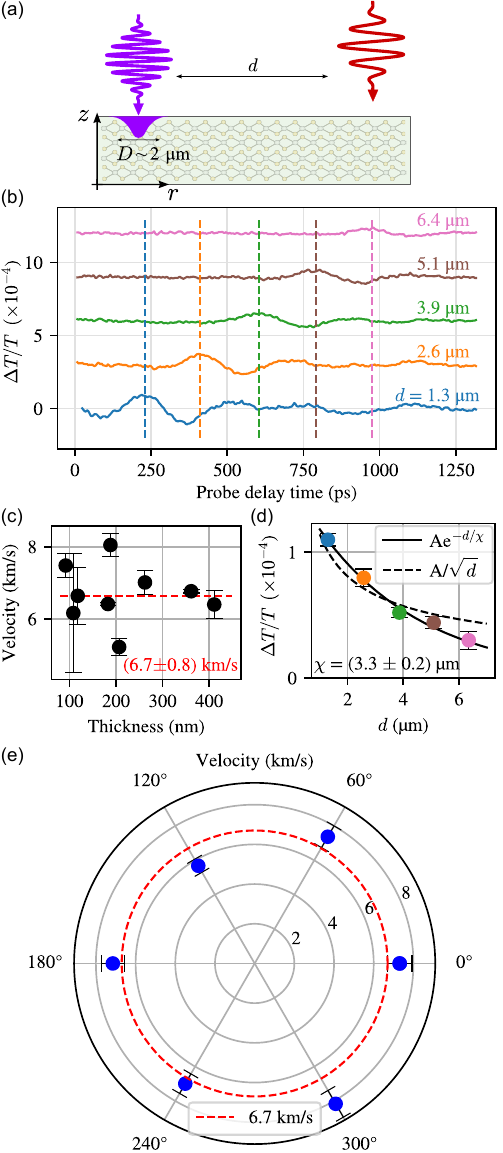}
  \caption{Propagation and attenuation of guided waves. (a) Schematic of the spatially decoupled pump–probe configuration: a 400 nm wavelength pump pulse excites acoustic vibrations, while an 800 nm wavelength probe pulse monitors the mechanical disturbance at a distance $d$ from the excitation area. (b) Representative spatially decoupled differential transmission signals for a 347~nm thick flake after thermal background subtraction, with $d$ ranging from 1.3~\textmu{}m to 6.4~\textmu{}m. The signals exhibit a wave packet arriving at increasingly later times. (c) Propagation velocity extracted from different sets of spatially decoupled measurements as a function of flake thickness. The dashed red line indicates the weighted average velocity of (6.7~$\pm$~0.8)~km~s$^{-1}$. (d) Amplitude of the first signal crest from panel (b) as a function of $d$. Solid and dashed black lines represent fits to exponential and geometric attenuation models, respectively, with the exponential fit yielding an attenuation coefficient $\chi$~=~(3.3~$\pm$~0.2)~\textmu{}m. (e) Velocity measurements performed at varying probe angles relative to the crystalline axes. The dashed red line represents the average velocity from panel (c), showing no significant in-plane anisotropy within experimental dispersion.}
  \label{fig2}
\end{figure}

MoS$_2$ flakes were mechanically exfoliated onto a flat silica substrate (see optical images of representative flakes in Figure~\ref{fig1}a). We measured the acoustic response of the flakes using an ultrafast optical pump–probe configuration (400~nm pump and 800~nm probe wavelengths; see the Experimental Section for details). Vibrational modes were studied using both the conventional in situ pump–probe technique (schematic in Figure~\ref{fig1}b, left) and a spatially decoupled variant, in which the pump and probe beams are spatially separated to probe the propagation of lattice-induced changes up to several micrometers away from the excitation region (schematic in Figure~\ref{fig2}a).

In situ differential transmission measurements yield a signal composed of multiple vibrational modes superimposed on a thermal relaxation background, which is subtracted using a multi-exponential fit. A representative signal for a flake with a thickness of (108~$\pm$~3)~nm is shown in Figure~\ref{fig1}b (top, black solid line), together with its Fast Fourier transform (FFT; bottom). The spectrum exhibits the fundamental breathing mode at 18.9~GHz, corresponding to a longitudinal out-of-plane vibration, along with a pronounced low-frequency peak at 2.7~GHz which we attribute to a guided plate-wave mode transporting acoustic energy radially outward from the optically excited region \cite{Zalalutdinov2021}. The flake thicknesses were determined by atomic force microscopy (AFM) and, where AFM was not performed, by exploiting the established relationship between thickness and the oscillation frequency of the fundamental breathing-mode, as reported elsewhere \cite{AVERSA2026}. 

As the flake thickness increases, the intensity of the low-frequency mode relative to the fundamental breathing mode increases. This trend is illustrated in Figure~\ref{fig1}c, which depicts the amplitude of the low-frequency mode normalized to that of the breathing mode for eight different flakes. In a previous study \cite{AVERSA2026} we demonstrated that the breathing mode undergoes three distinct energy dissipation mechanisms: surface scattering in thinner flakes \cite{Ziman1960,Cuffe2013}, substrate leakage for mid-range thicknesses \cite{Wright2023}, and conversion to guided wave modes for thicker flakes \cite{Zalalutdinov2021}. Consequently, as the thickness increases, the relative contribution of the other dissipation channels diminishes, leaving more energy available for conversion into guided wave modes, in agreement with the thickness-dependent increase observed in Figure~\ref{fig1}c.

Typical traces obtained from spatially decoupled pump-probe measurements are shown in Figure~\ref{fig2}b for a flake of thickness (347~$\pm$~5)~nm. A clear propagating wave is observed, whose amplitude decreases as the pump-probe separation is increased from 1.3 to 6.4~\textmu{}m. By tracking a fixed feature of the waveform (e.g., a crest) as a function of distance, assuming non-dispersive propagation, we extract a velocity of (6.77~$\pm$~0.05)~km~s$^{-1}$. Repeating this analysis for 10 flakes with thicknesses between 90 and 410~nm shows that the velocity is essentially thickness independent, with an average value of (6.7~$\pm$~0.8)~km~s$^{-1}$ (Figure~\ref{fig2}c). In addition, measurements performed while rotating the crystalline orientation reveal no detectable anisotropy in the propagation velocity within our experimental resolution (Figure~\ref{fig2}e). 

When the signal-to-noise ratio is sufficiently high, multiple successive crests can be resolved in a single trace, allowing for an independent determination of the wave frequency from the periodicity of the signal. For the traces shown in Figure~\ref{fig2}b, this procedure yields $f = $ (3.3 $\pm$ 0.8)~GHz, in excellent agreement with the in situ measurements performed on the same flake. Notably, the measured velocity is also consistent with a simple estimate based on the spatial extent of the excitation spot. The pump beam drives the sample over a finite region with an approximately Gaussian intensity profile, with a characteristic width of $\sim$2~\textmu{}m. This finite spot size determines the dominant in-plane wavevectors of the impulsive stress, so that the launched acoustic packet is expected to be composed primarily of wavelengths ($\lambda$) comparable to the excitation size, i.e., of a few micrometers. Combining this characteristic wavelength with the measured frequency range ($f \sim 2$--5~GHz) yields an expected acoustic velocity of order $v \sim f\lambda \sim 4$--10~km~s$^{-1}$, in good agreement with the velocities extracted directly from the time-of-flight analysis.

The acoustic energy attenuation with distance was evaluated by measuring the decay of the signal amplitude as a function of the pump-probe separation. Figure~\ref{fig2}d shows this dependence for the measurement presented in Figure~\ref{fig2}b, together with fits corresponding to exponential attenuation  ($\propto e^{-d/\chi}$) and to purely geometric cylindrical spreading ($\propto 1/\sqrt{d}$), where $d$ is the center-to-center separation of the pump and probe spots and $\chi$ the attenuation length. The better agreement with the exponential model indicates that the observed decay cannot be explained solely by geometric spreading in the plane of the flake, but instead reflects intrinsic dissipation mechanisms within the material. The exponential fit yields an attenuation length of $\chi$~=~(3.3~$\pm$~0.2)~\textmu{}m. This suggests that energy loss mechanisms beyond simple in-plane spreading contribute significantly to attenuation.

To extend our understanding of the guided wave acoustic modes exhibited by the MoS$_2$ flakes, we performed frequency-domain Finite Element Method (FEM) simulations (details provided in the Experimental Section). These simulations calculate atomic displacements within a continuum, isotropic, axially symmetric model. The input parameters for the simulations are the Young's modulus ($E$) and Poisson's ratio ($\nu$) of the materials, which were set to 225~GPa and 0.27, respectively. This parameter set was chosen because it reproduces the experimental response well, is consistent with prior reports \cite{Castellanos2012, Bertolazzi2011}, and yields longitudinal and shear wave velocities in agreement with both measurements \cite{Kim2017} and theoretical predictions \cite{Kaasbjerg2012}. In the simulation, the excitation following optical absorption
of the pump pulse is modeled via linear thermal expansion of the lattice.
The excitation is applied over a region defined by the pump spot, with a radius of 1.1~\textmu{}m ($e^{-2}$ intensity) and a penetration depth limited by ultrafast absorption \cite{Ge2014}. We evaluate the average absolute radial displacement, $|u_r|$, by integrating over the volume sampled by the probe beam. The probe radius is 0.8~\textmu{}m ($e^{-2}$ intensity), and because MoS$_2$ is transparent at 800~nm \cite{Beal1979}, its sampling depth extends through the entire flake thickness.

\begin{figure}
  \centering
  \includegraphics{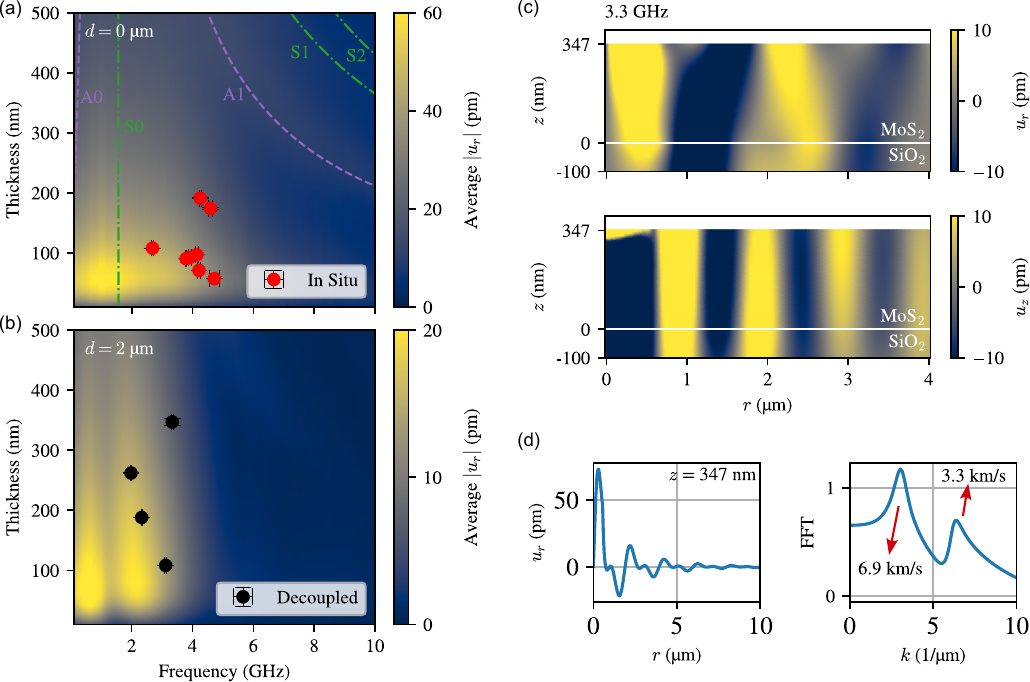}
  \caption{ FEM simulation results and mode analysis. (a, b) Colormaps of the simulated average absolute radial displacement ($|u_r|$) as a function of flake thickness and acoustic frequency for pump–probe distances $d$=~0~\textmu{}m and 2~\textmu{}m, respectively. Red and black markers indicate experimental data for in situ and decoupled measurements, respectively. Dashed lines in panel (a) represent theoretical thickness-to-frequency dispersion relationships for Lamb wave modes. (c) Simulated radial ($u_r$, top) and vertical ($u_z$, bottom) displacement profiles for a 347~nm thick MoS$_2$ film at a frequency of 3.3~GHz. The $z$-coordinate represents height, with $z$~=~0 at the SiO$_2$ substrate interface, while $r$ denotes the radial distance from the center of the excitation spot. (d) Radial displacement profile at the film surface ($z$~=~347~nm) as a function of $r$ (left) and its corresponding FFT (right). The spectral analysis reveals two distinct peaks corresponding to propagation velocities of 6.9~km~s$^{-1}$ and 3.3~km~s$^{-1}$, identifying the mixed longitudinal and shear nature of the radial signal.}
  \label{fig3}
\end{figure}

Figures~\ref{fig3}a and \ref{fig3}b present the simulated lattice response for the in situ and spatially separated configurations, respectively. Red and black markers indicate the corresponding experimental measurements, with each marker representing a different flake. For the spatially decoupled measurements, we include only those traces where the frequency can be independently determined, i.e., where at least two consecutive crests are clearly resolved. Due to strong attenuation, this condition is satisfied only at short pump-probe separations. For this reason, the corresponding simulations were performed at a representative separation of $d=2$~\textmu{}m.
The simulations predict a broad resonant region at low frequencies and small thicknesses, in good agreement with the experimental data for both configurations. The remaining dispersion is likely attributable to variations in the excitation conditions.

The in situ simulation graph (Figure~\ref{fig3}a) also includes dashed lines representing the theoretical dispersion relationships for the first two antisymmetric and the first three symmetric Lamb wave modes. These relations correspond to the classical solution for free-standing elastic plates and are commonly used as a reference framework for guided modes in thin films. Such modes are frequently discussed as the dominant channels for elastic energy leakage in thicker flakes \cite{Zalalutdinov2021, Soubelet2019}. Details of the Lamb-wave calculations, together with a saturated colormap used to identify these specific branches in the simulations, are provided in Supporting Information, Section~S2. As observed in Figure~\ref{fig3}a, the experimental data do not fall within the expected regions for Lamb wave modes. Figure~\ref{fig3}c illustrates the simulated initial $u_r$ and $u_z$ displacement profiles for a thickness of 347~nm at 3.3~GHz, matching a measurement data point. The radial displacement shows a mixture of different wavelengths, whereas the vertical displacement contains essentially a single harmonic component outside the pumped region. 

To further analyze the nature of these modes, we consider two points symmetrically located with respect to the plate's mid-plane. Specifically, in a reference system where $z$~=~0 is the bottom and $z$~=~$h$ is the surface of a flake with thickness h, we can select two points, $\mathbf{p}^{(1)}$ and $\mathbf{p}^{(2)}$, at a fixed radial position $r_0$ and vertical positions $z_1$ and $z_2$ (where $z_1$ $>$ $z_2$), located symmetrically relative to the center of the flake at $z$~=~$h$/2. 

In this framework, symmetric Lamb wave modes exhibit:
$$u^{(1)}_r = u^{(2)}_r$$
$$u^{(1)}_z = -u^{(2)}_z$$

\noindent where  $u^{(j)}_r$ and $u^{(j)}_z$ are the radial and vertical components of the displacement field at a point $\mathbf{p}^{(j)}$, respectively. Conversely, antisymmetric Lamb wave modes are characterized by:

$$u^{(1)}_r = -u^{(2)}_r$$
$$u^{(1)}_z = u^{(2)}_z$$

According to this symmetry convention, the displacement profile in Figure~\ref{fig3}c shows a predominantly symmetric $u_r$ component that progressively loses coherence as it propagates away from the excitation region, while $u_z$ exhibits antisymmetric behavior. This pattern is consistently observed across all simulations within this frequency-thickness range. These features indicate that the observed response is better described as a superposition of decoupled longitudinal and shear propagating modes, rather than as a single Lamb wave mode. In particular, the presence of the substrate modifies the boundary conditions and promotes mixing of the displacement components, especially in the radial response.

To test whether the deviation from the Lamb-wave description arises solely from substrate-modified boundary conditions, we performed complementary in situ measurements on suspended flakes (sample details in Supporting Information, Section~S1). The suspended-flake data (Supporting Information, Section~S2) show a similar mismatch with the Lamb-wave dispersion, indicating that the observed behavior reflects a general limitation of the ideal Lamb-wave description in this frequency-thickness regime. 

A profile of the radial displacement at the flake surface ($z$~=~347~nm) is shown in Figure~\ref{fig3}d (left) together with its FFT (right). This analysis reveals the presence of a fast longitudinal wave (6.9~km~s$^{-1}$), matching our experimental guided wave velocity, and a slow wave (3.3~km~s$^{-1}$), which is the sole component present in the vertical profile analysis of Figure~\ref{fig3}c. Supporting Information, Section~S2 includes additional simulation profiles and analyses of $u_r$ and $u_z$, along with higher-order Lamb-wave harmonics for reference. Time-domain simulations in Section~S3 confirm that the decoupled $u_r$ and $u_z$ components propagate with two distinct velocities. We note that, experimentally, only the fastest propagation velocity is observed, which can be understood from the optical detection mechanism. In optical pump-probe measurements on thin films, the differential transmission signal, $\Delta T/T$, contains a contribution that is proportional to the pump-induced modulation of the film thickness, $\Delta h$ \cite{Gusev1996, CARR2024}. As a result, vibrational modes that produce little or no out-of-plane thickness change generate a much weaker optical response and are effectively suppressed in $\Delta T/T$. This sensitivity mechanism, which accounts for the pronounced enhancement or suppression of breathing-mode harmonics depending on their parity, has been suggested elsewhere \cite{Schubert2015}. In these guided wave oscillations, longitudinal waves stretch or compress the material in the radial direction, which, in turn, causes the material to contract or expand in the perpendicular ($z$) direction. Through Poisson coupling, this in-plane strain induces a synchronous thickness modulation with the same periodicity as the longitudinal wave. In contrast, the antisymmetric shear modes observed in our simulations maintain a constant flake thickness throughout their propagation, rendering them largely invisible to the thickness-sensitive optical probe.

Finally, we present a strategy to efficiently excite guided-wave modes in flakes that couple only weakly to direct ultrafast optical pumping. Because thicker flakes exhibit a substantially stronger guided-mode response than thinner ones, selective excitation and detection can be achieved by pumping the thicker side of a flake and probing the thinner region.

\begin{figure}
  \centering
  \includegraphics{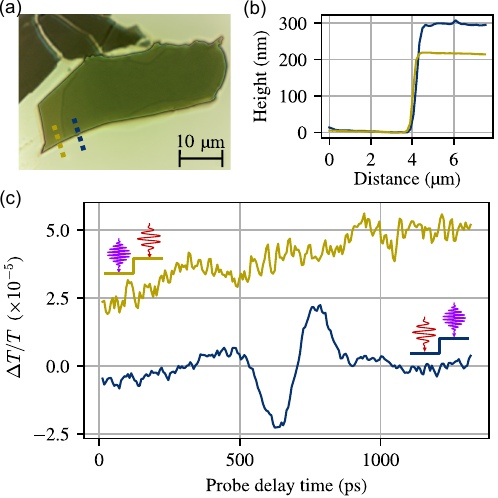}
  \caption{Guided wave transmission across a MoS$_2$ step. (a) Optical microscopy image of an exfoliated MoS$_2$ flake featuring two neighboring regions of different thicknesses that form a physical step. (b) AFM height profiles corresponding to the dashed lines in panel (a), revealing a step height of $\sim$80~nm (transitioning from 300~nm to 220~nm).(c) Spatially decoupled differential transmission signals measured at a pump–probe distance $d\sim$~4~\textmu{}m. The blue trace represents the signal obtained when the thicker region is pumped and the thinner region is probed, exhibiting successful wave transmission. The ocher trace shows the reciprocal configuration (pumping the thin region and probing the thick region), where no significant signal is detected above the noise floor.}
  \label{fig4}
\end{figure}

Figure~\ref{fig4}a shows an optical microscopy image of a flake comprising two thickness zones separated by a MoS$_2$ step. AFM profiles (Figure~\ref{fig4}b) yield thicknesses of 220~nm and 300~nm, corresponding to a step height of $\sim$80~nm. For this measurement, the pump power was set to 50~\textmu{}W, a regime in which a guided-wave signal is detectable only in the 300~nm thick region (see Supporting Information, Section~S4 for the signal amplitude dependence on pump power).

In this configuration, we pumped and probed at opposite ends of the step with a pump-probe distance of approximately 4~\textmu{}m. The signal obtained by pumping the thick end and probing the thin end is shown in blue in Figure~\ref{fig4}c, while the result of exchanging the pump and probe locations is shown in ocher. The first signal exhibits a pronounced dip arriving approximately 600~ps after excitation. In contrast, the second signal shows only variations on the order of the noise floor, which cannot be confidently attributed to this mode.

This demonstrates that when direct light-induced excitation of this guided wave mode is limited, the vibrational mode can be transmitted to the target flake by exciting a thicker, \textit{source} flake engineered to be in contact with the thinner one.

Simulations of the MoS$_2$ step were also performed (see Section S5 of the Supporting Information). Although the effects in the simulation are subtler than the experimental observations, integration of the displacement moduli over the probed volume yields a 1.5-fold enhancement in amplitude when the thicker part of the step is pumped. Given the short optical penetration depth at 400~nm wavelength, the total absorption should be identical in both configurations. Therefore, the discrepancy with the simulation may originate from enhanced optomechanical coupling when probing the thicker part of the flake.

\section{Conclusions}

In this work, we have comprehensively investigated the generation and propagation of guided acoustic waves in MoS$_2$ flakes using both in situ and spatially decoupled ultrafast pump–probe technique. 

By comparing experimental arrival times with finite-element-method (FEM) simulations, we identified a propagating acoustic wave with a constant velocity of (6.7~$\pm$~0.8)~km~s$^{-1}$, which remains constant across a thickness range of 90 to 410~nm, and exhibits frequencies in the 2–5~GHz range determined by the excitation profile.

Our findings reveal several critical insights into the nanomechanical behavior of TMDs. Spectral and symmetry analysis of the simulated displacement profiles suggest that the observed vibrations are not pure Lamb waves, but are better described as a superposition of decoupled longitudinal and shear propagating modes. We suggest that the Poisson effect governs the experimental optical readout of these modes. Longitudinal waves modulate the flake thickness ($\Delta h$), making them detectable via differential transmission ($\Delta T/T$), whereas antisymmetric shear modes maintain a nearly constant thickness and remain effectively \textit{invisible} to the probe.

We also established by attenuation measurements that the signal decay is dominated by intrinsic energy dissipation mechanisms rather than simple geometric spreading, with a characteristic attenuation coefficient $\chi \sim$~3.3~\textmu{}m.

Finally, we demonstrated a "source-target" engineering application where guided waves are successfully transmitted across a physical step in a flake. This allows for the excitation of acoustic modes in thin regions where direct optical excitation would otherwise be inefficient.

These results provide a fundamental framework for understanding how acoustic energy propagates laterally in two-dimensional semiconductors. By mastering the control of these guided modes, we open new pathways for the design of nanoscale phononic circuits, high-frequency optomechanical resonators, and hybrid quantum acoustic devices based on van der Waals heterostructures.

\section{Experimental Section}
\subsection{Sample preparation}
The SiO$_2$ substrates were cleaned by ultrasonicating them for 5 minutes each in acetone, isopropyl alcohol, and deionized water successively, followed by blow-drying with N$_2$ gas. MoS$_2$ flakes were mechanically exfoliated from bulk crystals using adhesive tape and transferred onto the cleaned SiO$_2$ substrates. AFM measurements were then performed on these flakes to determine their thicknesses accurately.

\subsection{Ultrafast pump probe measurement}
The output of a Ti:Sapphire laser—producing pulses of $\sim$50~fs duration, with a 95~MHz repetition rate, 800~nm wavelength and $\sim$300~mW average power—was focused onto a nonlinear crystal to generate the second harmonic at 400~nm. The second harmonic beam was modulated using an acousto-optic modulator ($\sim$100~kHz) to serve as the pump beam. The residual 800~nm beam was directed into a mechanical delay line with a 1.3~ns range and utilized as the probe in a transmission configuration.

At the sample, the pump average power was maintained at approximately 500~\textmu{}W for most measurements, while the probe power was set to 50~\textmu{}W. The spot radii ($e^{-2}$ intensity) were approximately 1.1~\textmu{}m for the pump and 0.8~\textmu{}m for the probe. All measurements were performed using lock-in detection. Differential transmission signals ($\Delta T/T$) were obtained by normalizing the transient signal to direct transmission measurements.

Both spatially overlapped (in situ) and spatially decoupled pump-probe measurements were performed on approximately 20 different MoS$_2$ flakes. To isolate the acoustic contribution, the thermal background was removed using multi-exponential or linear fits. The resulting residual signals were either fitted with damped sine functions (for in situ measurements) or analyzed statistically (for decoupled measurements).

\subsection{FEM Simulations}

Frequency-domain numerical calculations were performed using the Structural Mechanics module of COMSOL Multiphysics. A 2D $(r, z)$ domain with axial symmetry was implemented, using perfectly matched layer (PML) boundary conditions to truncate the computational domain and simulate a semi-infinite substrate. Simulations were conducted for MoS$_2$ thicknesses ranging from 10~nm to 1000~nm atop a SiO$_2$ substrate, surrounded by air.

The material properties used for MoS$_2$ were a Young's modulus ($E$) of 225~GPa and a Poisson's ratio ($\nu$) of 0.27. For the SiO$_2$ substrate, the values used were 73~GPa and 0.17, respectively\cite{FUKUHARA1997}. A thermal strain $\varepsilon=\alpha (\Theta - \Theta_{\textrm{ref}})$, where $\alpha$ is the linear thermal expansion coefficient (8$\times$10$^{-6}$ K$^{-1}$) and $\Theta-\Theta_{ref}=100$ K, was implemented to simulate the pump excitation, with a spot radius of 1.1~\textmu{}m ($e^{-2}$ intensity) and a penetration depth of 75~nm. The linear elastic response ($u_r$ and $u_z$) of the films was obtained by solving Navier’s equation in the frequency domain from 0.1~GHz to 10~GHz. To simulate the experimental response, the displacement was integrated over the volume defined by the probe beam (radius of 0.8~\textmu{}m and depth equal to the flake thickness) at varying distances $d$ from the excitation source.

To account for intrinsic damping, an isotropic loss factor ($\eta$) of 0.113~\textmu{}m was implemented in the MoS$_2$ domain, corresponding to the experimentally observed attenuation length ($\chi$) of 3.3~\textmu{}m. Continuity of the stress and displacement fields was enforced at all internal boundaries.

A complementary time-domain simulation was performed from 0 to 1400~ps for a MoS$_2$ flake with a thickness of 347~nm (see Supporting Information, section S3). This simulation was designed to validate the frequency-domain results and confirm the propagation velocities obtained experimentally.

\medskip

\medskip
\textbf{Acknowledgements} \par 
This work was partially supported by ANPCyT (PICT 2021 IA-363 and PICT 2021 GRF-TI-349; no disbursements since December 2023), CONICET (PIP 112-20200101465), UBACyT (Proyecto 20020220200078BA), and Red Federal de Alto Impacto (FFFLASH; CONVE-2023-10189190). A.V.B. acknowledges funding support from the Alexander von Humboldt Foundation through the 2023 Georg Forster Award.
The authors thank Dr. I. Abdelwahab (Harvard University) for providing the bulk crystals used for exfoliation.
\medskip

\bibliographystyle{unsrt} 

\bibliography{References}

\clearpage 
\appendix

\titleformat{\section}{\normalfont\Large\bfseries\centering}{}{0pt}{}

\setcounter{subsection}{0}
\renewcommand{\thesubsection}{S\arabic{subsection}}

\setcounter{figure}{0}
\renewcommand{\thefigure}{S\arabic{figure}}
\setcounter{table}{0}
\renewcommand{\thetable}{S\arabic{table}}

\section{Supporting Information}
\label{appendix:SI}

\subsection{Suspended flakes}

In addition to measurements conducted on MoS$_2$ flakes supported on flat SiO$_2$ substrates, further investigations were performed on flakes decoupled from substrate influence—specifically, in suspended configurations. To achieve this, borosilicate glass substrates were patterned with an array of groove-shaped micro-cavities using a photolithography process.

\begin{figure}[h]
  \centering
  \includegraphics{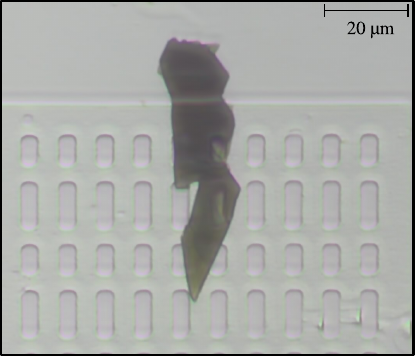}
  \caption{Optical microscopy image of mechanically exfoliated MoS$_2$ flakes on a groove-shaped holes patterned borosilicate substrate.}
  \label{si:fig1}
\end{figure}

Figure~\ref{si:fig1} presents an optical microscopy image of a flake transferred onto such a substrate, showing the regions spanning the patterned grooves. These suspended configurations allow for the characterization of intrinsic acoustic properties by eliminating interfacial damping and mechanical coupling with the underlying glass.

We successfully performed four in situ measurements on suspended flakes; however, the dimensions of the suspended regions did not allow for spatially decoupled measurements across several micrometers.

\begin{figure}[h]
  \centering
  \includegraphics{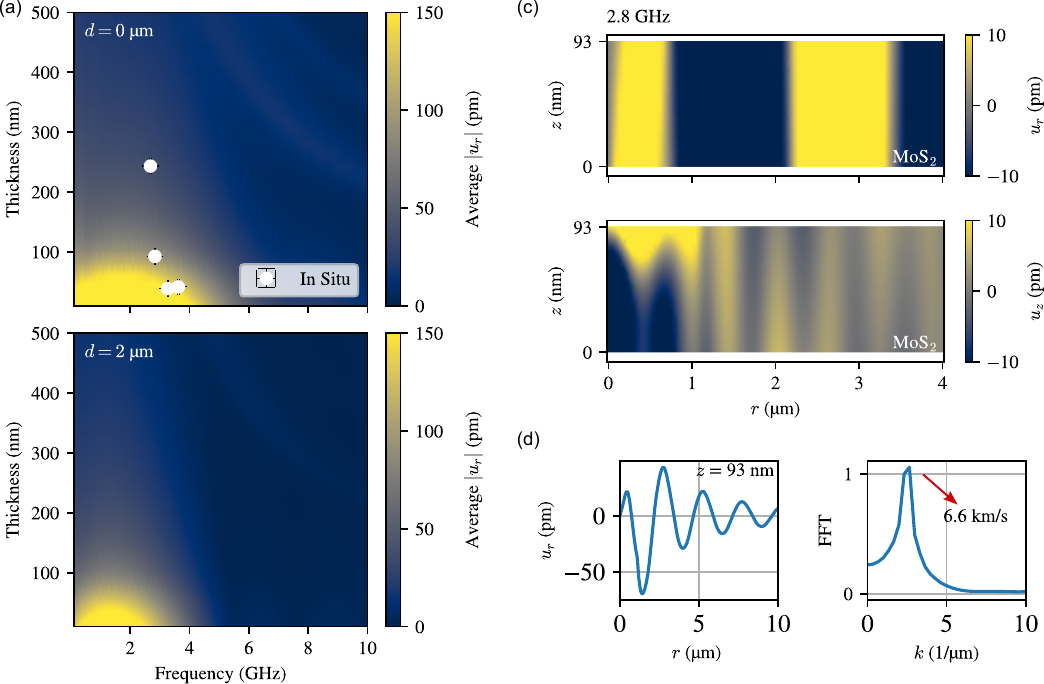}
  \caption{(a,b) Colormaps of the simulated average absolute radial displacement ($|u_r|$) as a function of flake thickness and frequency for pump–probe distances $d$~=~0~\textmu{}m and 2~\textmu{}m, respectively, in a suspended configuration. White markers indicate experimental data for in situ measurements. (c) Simulated radial ($u_r$, top) and vertical ($u_z$, bottom) displacement profiles for a 93~nm thick MoS$_2$ film at a frequency of 2.8~GHz. The $z$-coordinate represents height, with $z$~=~0 the bottom surface and $z$~=~93~nm the top surface. (d) Radial displacement profile at the film surface ($z$~=~93~nm) as a function of $r$ (left) and its corresponding FFT (right). The spectral analysis reveals only one peak corresponding to a propagation velocity of 6.6~km/s.}
  \label{si:fig2}
\end{figure}

Figure~\ref{si:fig2}a shows the results of these measurements (white markers) overlaid on the colormap of the simulated average radial displacement $|u_r|$, integrated at zero distance from the excitation region. This simulation assumes a MoS$_2$ film surrounded by air both above and below to represent the suspended configuration. As in Fig. 3a of the main manuscript (supported flakes), the experimental data lie reasonably within the resonant region of the colormap.

In Figure~\ref{si:fig2}b, the average $|u_r|$ is shown integrated for a pump–probe separation of 2~\textmu{}m. Here, the resonant region shifts to the red relative to the in situ colormap. Figure~\ref{si:fig2}c displays the radial and vertical displacement profiles for a 2.8~GHz oscillation in a 93~nm thick film (matching an experimental data point). These profiles exhibit the characteristic decoupled longitudinal and shear propagation modes, respectively. Furthermore, symmetry analysis identifies a symmetric mode in the radial profile, while the vertical profile outside the pumped region exhibits an antisymmetric character.

Finally, Figure~\ref{si:fig2}d shows the one-dimensional radial displacement at the surface of the flake and its corresponding FFT. This spectral analysis reveals that, in the suspended configuration, there is no mixing of wave modes in the radial direction, contrasting with the behavior observed in the supported flakes discussed in the main manuscript. This pure longitudinal mode propagates at a velocity of 6.6~km/s.

\subsection{Lamb waves}

\begin{figure}[h]
  \centering
  \includegraphics{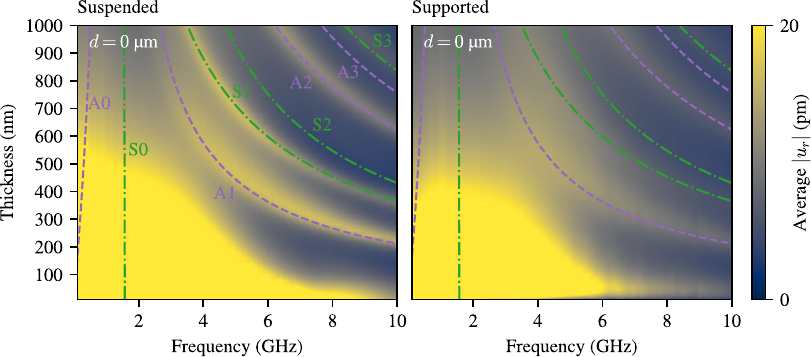}
  \caption{Colormaps of the simulated average absolute radial displacement ($|u_r|$) as a function of flake thickness and frequency for a pump–probe distance of $d$~=~0~\textmu{}m for suspended (left) and supported (right) configurations. Dashed lines represent theoretical thickness-to-frequency dispersion relationships for Lamb wave modes. Resonant regions in the colormap coincide with modes A1, S1, A2 and S3.}
  \label{si:fig3}
\end{figure}

In this section, we demonstrate that the simulation results accurately capture the fundamental Lamb wave modes. Figure~\ref{si:fig3} presents colormaps of the average radial displacement absolute value ($|u_r|$) for both suspended (left) and supported (right) configurations at $d$~=~0~\textmu{}m, utilizing a saturated color scale to highlight lower-intensity features.

In these plots, the dashed lines represent theoretical calculations of the Lamb wave dispersion relationships. These analytical curves coincide with the resonant regions in the colormaps for both configurations. Notably, while the agreement is clear in both cases, the resonant branches in the supported configuration appear more diffuse probably due to the mechanical coupling and energy leakage into the underlying substrate.

\begin{figure}[h]
  \centering
  \includegraphics{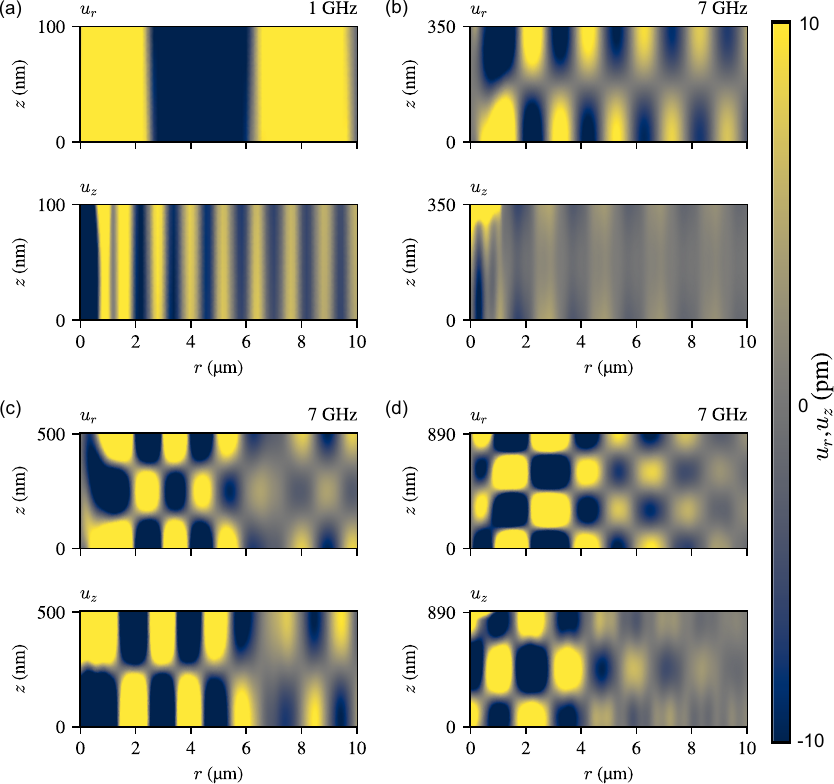}
  \caption{Simulated radial ($u_r$, top) and vertical ($u_z$, bottom) displacement profiles for suspended MoS$_2$ films. Panel (a) shows the profiles for the experimental regime at (1~GHz,~100~nm), while panels (b), (c), and (d) correspond to higher-order modes at (7~GHz,~350~nm), (7~GHz,~500~nm), and (7~GHz,~890~nm), respectively. The $z$-coordinate represents the film height (thickness), and $r$ represents the radial distance from the excitation source.}
  \label{si:fig4}
\end{figure}

These functions are derived from the Cauchy-Navier equation:

$$\mu \nabla^{2}\mathbf{u} + (\lambda + \mu)\nabla(\nabla \cdot \mathbf{u}) + \mathbf{f} = \rho \frac{\partial^{2}\mathbf{u}}{\partial t^{2}}$$

\noindent where $ \mu$ and $\lambda$ are the Lamé parameters, $\mathbf{u}$ represents the lattice displacement (treated as a continuum), $\mathbf{f}$ denotes the body forces, and $\rho$ is the material mass density.

This equation can be solved using the Helmholtz decomposition method, where the displacement field is decomposed into an irrotational part and a solenoidal part: $\mathbf{u} = \nabla \phi + \nabla \times \mathbf{\psi}$. For plane strain conditions (in $xz$ coordinates), this method yields two uncoupled wave equations governing longitudinal and transverse (shear) waves:

$$
\frac{\partial^2\phi}{\partial x^2}+\frac{\partial^2\phi}{\partial z^2}=\frac{1}{c_L^2}\frac{\partial^2\phi}{\partial t^2}
$$

$$
\frac{\partial^2\psi}{\partial x^2}+\frac{\partial^2\psi}{\partial z^2}=\frac{1}{c_T^2}\frac{\partial^2\psi}{\partial t^2},
$$

\noindent where $c_L$ and $c_T$ are the acoustic velocities of the longitudinal and transverse waves in the material, respectively. These velocities are related to the Lamé parameters as follows:

$$
c_{L}=\sqrt{\frac{\lambda +2\mu }{\rho }}
$$

$$
c_T=\sqrt{\frac{\mu}{\rho}}.
$$

An alternative set of parameters useful in this description are the Young's modulus ($E$) and Poisson's ratio ($\nu$), which are related to the Lamé parameters by the following equations:

$$
E=\frac{\mu (3\lambda +2\mu )}{\lambda +\mu }
$$
$$
\nu =\frac{\lambda }{2(\lambda +\mu )}.
$$

These equations must be solved subject to stress-free boundary conditions at the top and bottom surfaces of a film of thickness $h$. By proposing plane-wave solutions for the potentials $\phi$ and $\psi$, as detailed by Giurgiutiu (2014)\cite{Giurgiutiu2014}, the algebraic derivation yields two decoupled transcendental equations. These equations describe the symmetric and antisymmetric modes with respect to the mid-plane of the film, known as the Rayleigh-Lamb equations:

$$
\frac{\tanh (\eta_L h/2)}{\tanh (\eta_T h/2)}=-\frac{(k ^{2}-\eta_T ^{2})^{2}}{4\eta_T \eta_L k ^{2}} \textrm{, for symmetric modes and}
$$

$$
\frac{\tanh (\eta_L h/2)}{\tanh (\eta_T h/2)}=-\frac{4\eta_T \eta_L k ^{2}}{(k ^{2}-\eta_T ^{2})^{2}}\textrm{, for antisymmetric modes,}
$$

\noindent where

$$
\eta_L^2=\frac{\omega^2}{c_L^2}-k^2 \textrm{,}
$$

$$
\eta_T^2=\frac{\omega^2}{c_T^2}-k^2,
$$

\noindent where $\omega$ and $k$ are the angular frequency and the wavenumber of the acoustic wave, respectively.

This set of transcendental equations must be solved numerically to obtain the dispersion relationships. The resulting solutions correspond to the dashed lines overlaid on the simulation colormaps in Figure~\ref{si:fig3}, providing a direct comparison between the analytical Rayleigh-Lamb theory and our finite element model.

The simulation results clearly resolve the Lamb wave modes A1, S1, A2, and S3, whereas other modes within this spectral range do not appear in the simulation, likely due to the specific symmetry of the thermal excitation source. For a more detailed comparison of the acoustic modes in the experimental regime with analytical Lamb wave theory, Figure~\ref{si:fig4} presents the simulated radial and vertical displacement profiles for suspended films at the specific coordinates of (1~GHz,~100~nm) in panel (a), representing a region close to the experimental data, as well as (7~GHz,~350~nm), (7~GHz,~500~nm), and (7~GHz,~890~nm) in panels (b), (c), and (d), which correspond to the A1, S1, and A2 modes, respectively.

In panels (b), (c), and (d), the same symmetry for $u_r$ and $u_z$ can be observed across the modes. In contrast, the profiles in panel (a) exhibit opposing symmetries, highlighting that the vibrational character in the experimental regime differs from the higher-order Lamb branches identified at higher frequencies and thicknesses.

\begin{figure}[h]
  \centering
  \includegraphics{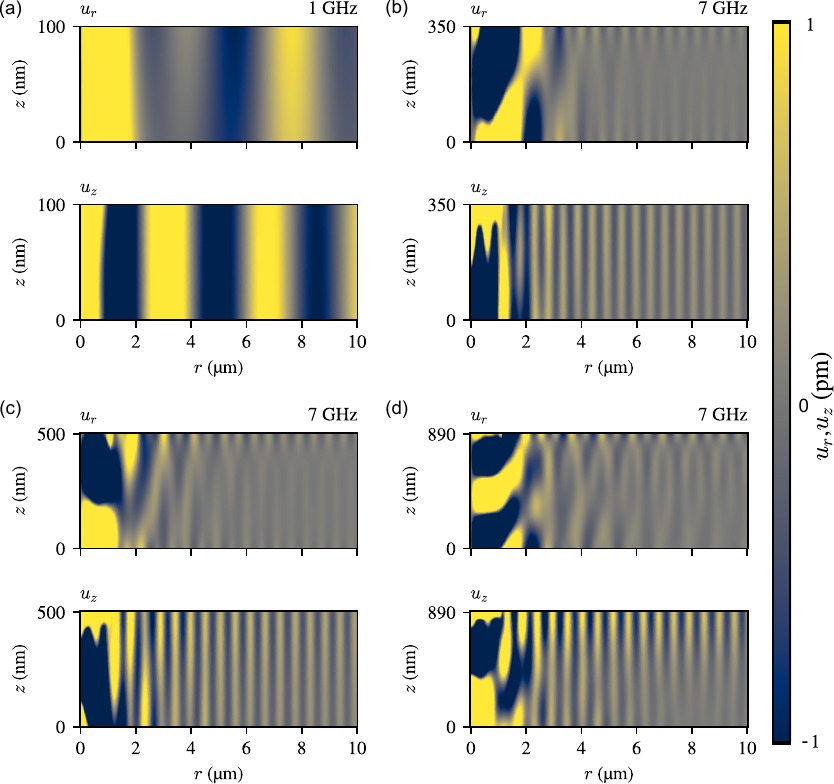}
  \caption{Simulated radial ($u_r$, top) and vertical ($u_z$, bottom) displacement profiles for supported MoS$_2$ films. Panel (a) shows the profiles for the experimental regime at (1~GHz,~100~nm), while panels (b), (c), and (d) correspond to higher-order modes at (7~GHz,~350~nm), (7~GHz,~500~nm), and (7~GHz,~890~nm), respectively. The $z$-coordinate represents the film height (thickness), and $r$ represents the radial distance from the excitation source.}
  \label{si:fig5}
\end{figure}

Figure~\ref{si:fig5} presents the same simulation parameters shown in Figure~\ref{si:fig4}, but for MoS$_2$ films supported on a SiO$_2$ substrate. In this configuration, while the analytical Lamb wave dispersion relationships still coincide with the resonant regions in Figure~\ref{si:fig3}, the symmetry analysis becomes more complex. The presence of the substrate relaxes the stress-free boundary conditions at the bottom interface, leading to coupled equations that facilitate mode mixing. Consequently, the well-defined symmetries characteristic of suspended films are often diminished or lost. Nevertheless, the simulation in panel (a) of this figure reveals a clearly symmetric $u_r$ profile and an antisymmetric $u_z$ profile. This observation provides strong evidence for our interpretation that, in this experimental regime, we are observing decoupled longitudinal and transverse oscillation modes rather than the higher-order coupled Lamb branches.

\begin{figure}[h]
  \centering
  \includegraphics{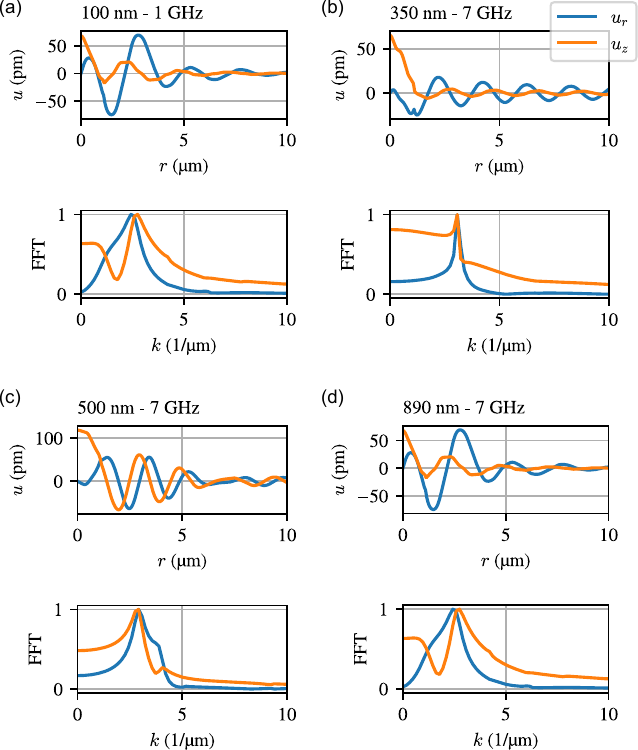}
  \caption{Radial displacement profile at the film surface as a function of $r$ (top) and its corresponding FFT (bottom) for suspended MoS$_2$ films. Panel (a) shows the profiles for the experimental regime at (1~GHz,~100~nm), while panels (b), (c), and (d) correspond to higher-order modes at (7~GHz,~350~nm), (7~GHz,~500~nm), and (7~GHz,~890~nm), respectively.}
  \label{si:fig6}
\end{figure}

\begin{figure}[h]
  \centering
  \includegraphics{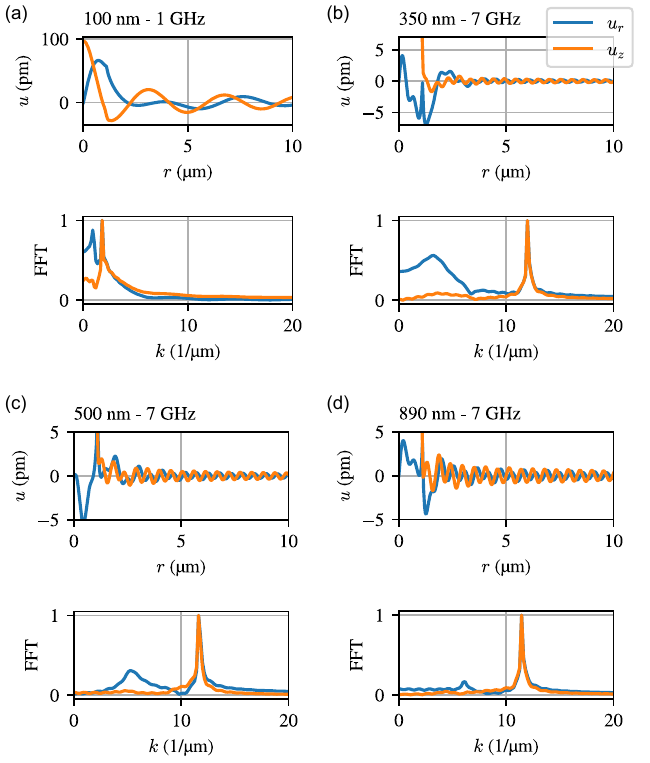}
  \caption{Radial displacement profile at the film surface as a function of $r$ (top) and its corresponding FFT (bottom) for supported MoS$_2$ films. Panel (a) shows the profiles for the experimental regime at (1~GHz,~100~nm), while panels (b), (c), and (d) correspond to higher-order modes at (7~GHz,~350~nm), (7~GHz,~500~nm), and (7~GHz,~890~nm), respectively.}
  \label{si:fig7}
\end{figure}

Furthermore, we can analyze the propagation velocity of these oscillation modes through their spatial profiles. One-dimensional radial and vertical displacement profiles, extracted from the film surface, are presented in Figure~\ref{si:fig6} for suspended films and Figure~\ref{si:fig7} for supported films, each accompanied by its corresponding FFT.

Following the same coordinate convention as before, panel (a) corresponds to the experimental regime at (1~GHz,~100~nm), while panels (b), (c), and (d) correspond to the A1, S1, and A2 modes at (7~GHz,~350~nm), (7~GHz,~500~nm), and (7~GHz,~890~nm), respectively. For the supported films, the FFTs are calculated starting from $r$~=~2~\textmu{}m; this spatial filtering is necessary to clear the spectra of the low wavenumber components and thermal artifacts that are superimposed within the immediate excitation area.

The analysis of the harmonic components in the suspended configurations reveals that for coordinates corresponding to higher-order Lamb wave modes, the propagation velocities extracted from both radial and vertical displacements are roughly identical. In contrast, at the 1~GHz and 100~nm coordinate, the radial and vertical velocities are clearly distinct, further supporting the decoupled nature of the acoustic modes in this regime. These extracted velocity values are summarized in Table~\ref{tab:velsuspended}.

\begin{table}[h]
    \centering
    \begin{tabular}{c||c|c|c|c}
    Simulated Point &(1~GHz,~100~nm)& (7~GHz,~350~nm)& (7~GHz,~500~nm)  & (7~GHz,~890~nm) \\ \hline \hline
    Radial velocity (km/s)   & 7.2 & 14.3 & 15.1& 18.1\\
    Vertical velocity (km/s)  & 1.1 & 14.3 & 15.1 & 16.0

    \end{tabular}
    \caption{Calculated velocities for the waves in the radial and vertical profiles in the suspended configuration (Fig.~\ref{si:fig6}). }
    \label{tab:velsuspended}
\end{table}

On the other hand, the harmonic components of the displacement profiles in the supported configurations reveal that the vertical displacement consistently exhibits a slow velocity component across all cases.

In contrast, the radial displacement shows a mixing of this slow wave mode with a higher velocity component. This hybrid character in the radial signal further illustrates the mode coupling induced by the substrate interface. These extracted velocity values are summarized in Table \ref{tab:velsupported}.

\begin{table}[h]
    \centering
    \begin{tabular}{c||c|c|c|c}
    Simulated Point &(1~GHz,~100~nm)& (7~GHz,~350~nm)& (7~GHz,~500~nm)  & (7~GHz,~890~nm) \\ \hline \hline
    Radial velocities (km/s)   & 3.3/6.6 & 3.7/13.7 & 3.8/8.5& 3.9/7.2\\
    Vertical velocity (km/s)  & 3.3 & 3.7 & 3.8 & 3.9

    \end{tabular}
    \caption{Calculated velocities for the waves in the radial and vertical profiles in the supported configuration (Fig.~\ref{si:fig7}). }
    \label{tab:velsupported}
\end{table}

\clearpage

\subsection{Time domain simulations}

\begin{figure}[h]
  \centering
  \includegraphics{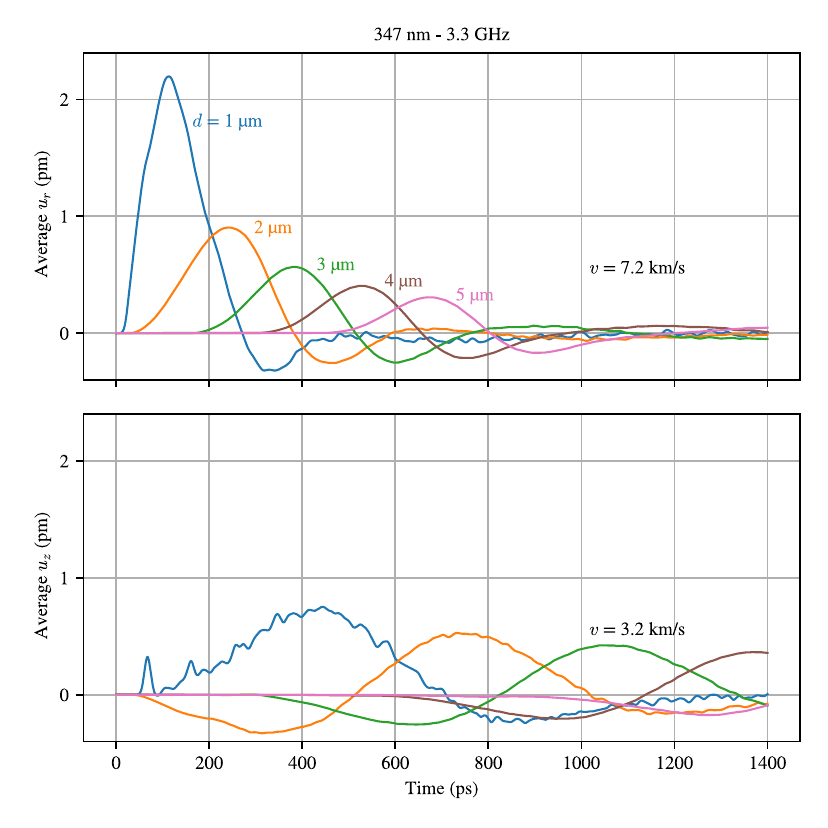}
  \caption{Time-domain simulations of acoustic wave propagation. The panels show the average radial displacement ($u_r$, top) and average vertical displacement ($u_z$, bottom) integrated over the probe beam volume as a function of time, for a 347~nm-thick supported flake  at 3.3~GHz. Data are presented for pump–probe distances $d$ of 1~\textmu{}m (blue), 2~\textmu{}m (orange), 3~\textmu{}m (green), 4~\textmu{}m (brown), and 5~\textmu{}m (pink). The calculated propagation velocities are 7.2~km/s and 3.2~km/s for the longitudinal (radial) and transverse (vertical) modes, respectively.}
  \label{si:fig8}
\end{figure}

To further validate the propagation velocities of the radial and vertical waves identified in the frequency-domain analysis, Fig.~\ref{si:fig8} shows complementary time-domain simulations for a 347~nm-thick supported film excited at 3.3~GHz. The top panel displays the average radial displacement $|u_r|$, while the bottom panel displays the average vertical displacement $|u_z|$, integrated in the volume defined by the probe beam for pump–probe distances $d$ ranging from 1~\textmu{}m to 5~\textmu{}m.

These simulations resolve wave packets propagating at a longitudinal velocity of 7.2~km/s and a transverse velocity of 3.2~km/s, respectively, in agreement with the frequency-domain analysis within 5$\%$.

\subsection{Power dependence}

\begin{figure}[h]
  \centering
  \includegraphics{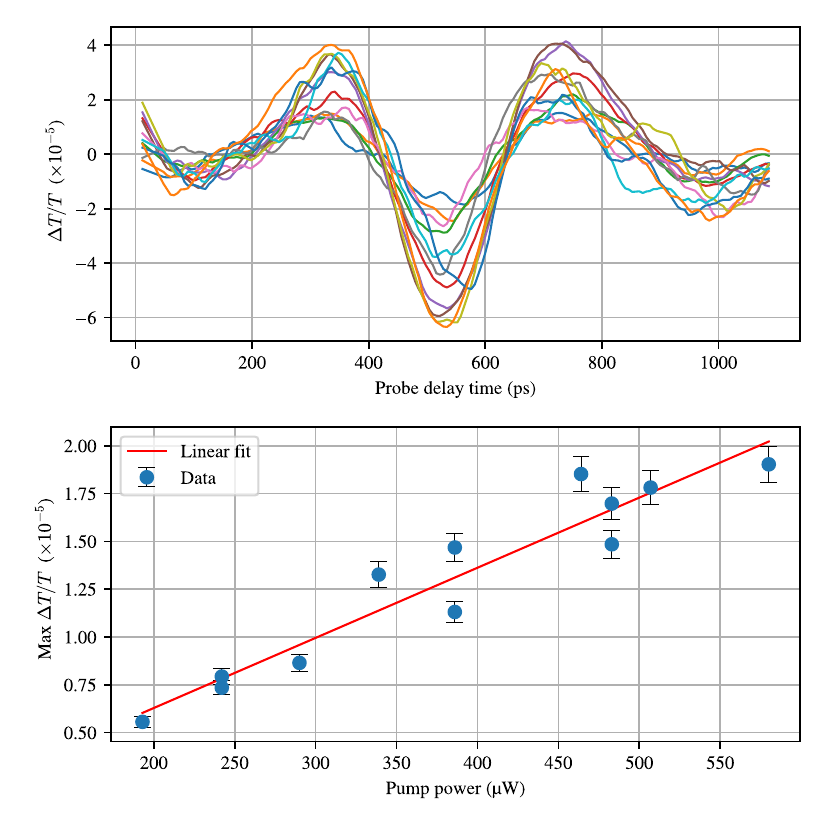}
  \caption{Power dependence of the acoustic signal. The top panel shows the differential signals ($\Delta T/T$) measured by varying the pump power for a fixed pump–probe distance of 3.9~\textmu{}m. The bottom panel shows the maximum amplitude of the $\Delta T/T$ signal as a function of the pump power (blue markers), accompanied by a linear fit (red solid line) that confirms the linear thermoelastic regime of the acoustic excitation.}
  \label{si:fig9}
\end{figure}

Pump power is a critical parameter for exciting detectable guided acoustic waves in this experimental configuration. The top panel of Figure~\ref{si:fig9} displays the differential signal measurements for an MoS$_2$ flake with a fixed pump–probe separation of 3.9~\textmu{}m at increasing pump powers.

The bottom panel presents the maximum signal amplitude obtained in each measurement as a function of the pump power. The data exhibit a clear linear trend within the range of 190 to 570~\textmu{}W, indicating that the acoustic excitation remains within the linear regime.

\subsection{Step Simulations}
\begin{figure}[h]
  \centering
  \includegraphics{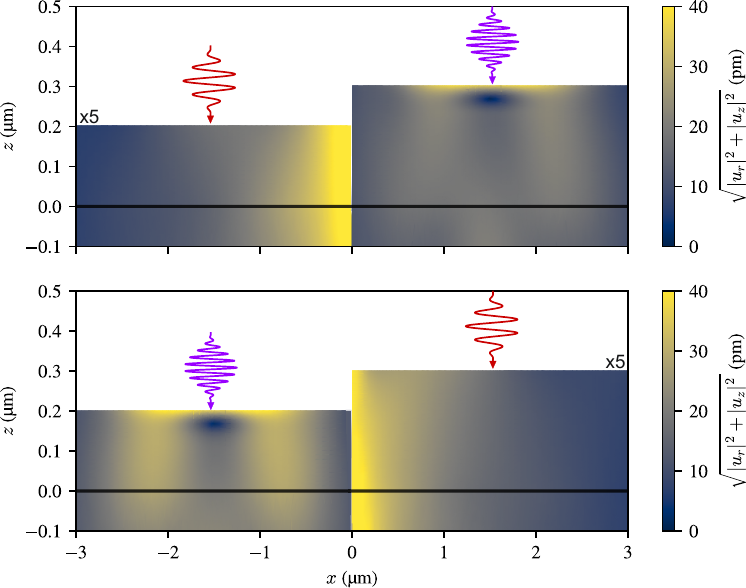}
  \caption{Simulated total displacement profiles for a configuration of MoS$_2$ films of differing thicknesses forming a step-edge. The top panel depicts the configuration where the thicker flake is pumped and the thinner flake is probed. The bottom panel shows the inverse configuration, where the beam locations are reversed. In both cases, the displacement magnitude in the probed region is saturated by a factor of 5 to enhance the visibility of the transmitted acoustic energy.}
  \label{si:fig10}
\end{figure}

Figure~\ref{si:fig10} shows the total displacement profiles for the step-edge simulation presented in the main manuscript. The top panel depicts the configuration where the thicker section of the flake is pumped and the thinner section is probed, while the bottom panel shows the inverted configuration. In both cases, the displacement in the probed region is saturated by a factor of 5 to enhance the visualization of subtle effects.

When the thinner flake is probed (top panel), the colormap reveals that energy transmission occurs homogeneously across the entire thickness ($z$-axis). Conversely, in the bottom panel—where the thicker flake is probed—the energy density narrows in the portion of the thick flake that extends beyond the contact area with the thin flake. This redistribution of energy results in a total integrated displacement in the probe volume of the top configuration that is 1.5 times greater than the integrated magnitude in the bottom configuration.

\end{document}